\documentclass[a4paper,12pt]{article}
\usepackage{a4wide}
\usepackage{graphicx}
\usepackage{epstopdf}
\usepackage{epsfig}
\usepackage[sumlimits,intlimits,namelimits]{amsmath}
\usepackage{amssymb,color}
\usepackage[english]{babel}

\numberwithin{equation}{section}
\usepackage{cite}

\begin{document}
\begin{titlepage}
\begin{flushright}
SI-HEP-2008-10 \\[0.2cm]
CERN-PH-TH/2008-117 \\[0.2cm]
\today
\end{flushright}

\vspace{1.2cm}
\begin{center}
\Large\bf\boldmath
Neutrino-Mass Hierarchies and \\
Non-linear Representation of \\
Lepton-Flavour Symmetry 

\unboldmath
 \end{center}

\vspace{0.5cm}
\begin{center}
{\sc Thorsten Feldmann$\,^a$ and Thomas Mannel$\,^{a,\,b}$} 

\vspace{1.4em}

{\small ${}^a$ {\sf Theoretische Physik 1, Fachbereich Physik,
Universit\"at Siegen, D-57068 Siegen, Germany.}
\vspace{0.7em}

${}^b$ {\sf CERN, Department of Physics, Theory Unit,  CH-1211 Geneva 23,
Switzerland.}}

\end{center}

\vspace{3em}
\begin{abstract}
Lepton-flavour symmetry in the Standard Model is broken
by small masses for charged leptons and neutrinos. Introducing
neutrino masses via dimension-5 operators associated to 
lepton-number violation at a very high scale, the corresponding
coupling matrix may still have entries of order 1, resembling
the situation in the quark sector with large top Yukawa coupling.
As we have shown recently, in such a situation one may introduce
the coupling matrices between lepton and Higgs 
fields as non-linear representations of 
lepton-flavour symmetry within an effective-theory
framework. This allows us to separate the effects 
related to the large mass difference observed in atmospheric neutrino 
oscillations from those related to the solar mass difference.
We discuss the cases of normal or inverted hierarchical and
almost degenerate neutrino spectrum, give some examples to illustrate 
minimal lepton-flavour violation in radiative and leptonic decays,
and also provide a systematic definition of next-to-minimal lepton-flavour
violation within the non-linear framework.

\end{abstract}

\end{titlepage}

\newpage
\pagenumbering{arabic}

\section{Introduction}

The gauge sector of the Standard Model (SM) is symmetric under
independent unitary transformations between the three family members
of each fermion multiplet (left-handed quarks and leptons, right-handed
up-, down-quarks and charged leptons). The Yukawa couplings between fermion
fields and the scalar Higgs field break the flavour symmetry, giving rise
to fermion masses and quark mixing. 

New Physics (NP) models generically introduce new sources of flavour symmetry
breaking, which are already highly constrained by precision data on B-meson
and kaon decays. To account for this observation, the concept of 
minimal flavour violation (MFV) has been proposed, which can be 
introduced in an elegant way by considering the Yukawa matrices of the
SM as vacuum expectation values (VEVs) of spurion fields
\cite{D'Ambrosio:2002ex,Cirigliano:2005ck} (for earlier, phenomenological definitions
of MFV, see also \cite{Ciuchini:1998xy,Buras:2000dm}).
NP effects can then be encoded in terms of higher-dimensional operators in an effective
theory (ET), where all flavour coefficients are proportional to SM masses and mixing parameters.

In a recent paper \cite{Feldmann:2008ja}, 
we have pointed out the particular role of the
top quark in the ET construction. 
Being the only fermion in the SM with Yukawa couplings
of order 1, the top quark breaks the flavour symmetry already at the cut-off scale $\Lambda$
of the ET. Therefore it is preferable to represent flavour symmetry
in a non-linear way in terms of Goldstone modes for broken flavour symmetry generators
and spurion fields which transform under the residual symmetry.

At first glance, the lepton sector in the SM does not contain large Yukawa couplings,
and therefore the usual (linear) representation of lepton flavour symmetry
could be applied to introduce
minimal lepton flavour violation (MLFV) \cite{Cirigliano:2005ck}. 
However, in a scenario with minimal field content 
(i.e.\ potential right-handed neutrinos having been integrated out),
the observed small neutrino masses have to be generated by
higher-dimensional lepton-number (LN) violating operators (see below). The small size of neutrino masses is naturally explained by the large scale $\Lambda_{\rm LN}$ associated to LN violation, while some of
the flavour coefficients of LN-violating operators 
(related to the largest eigenvalue in the neutrino mass matrix) 
may still be of order 1.

\begin{figure}
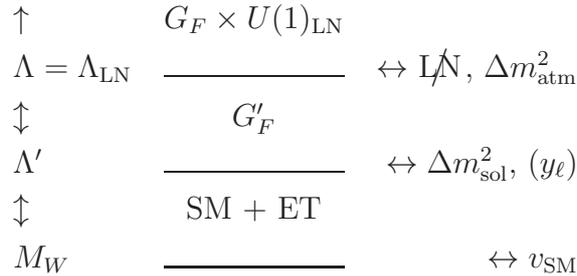

\begin{center}
\begin{tabular}{l c  r}
  $\uparrow$ &   $G_F \times U(1)_{\rm LN}$ & \\[0.25em]
  $\Lambda=\Lambda_{\rm LN}$ & \multicolumn{1}{c}{\hrulefill} 
   & $\leftrightarrow$ LN\hspace{-1em}{\large /}\hspace{0.5em},
       $\Delta m^2_{\rm atm}$ \\[0.35em]
  $\updownarrow$ & $G_F'$ & \\[0.25em]
  $\Lambda'$ & \multicolumn{1}{c}{\hrulefill} 
& $\leftrightarrow \Delta m^2_{\rm sol}$, ($y_{\ell}$) \\[0.25em]
  $\updownarrow$ & SM + ET & \\[0.35em]
  $M_W$ & \multicolumn{1}{c}{\hrulefill} &  $\leftrightarrow v_{\rm SM}$
  \end{tabular}
\vspace{0.8em}
\parbox{0.8\textwidth}{\small
\caption{ \label{illustration} \small Tower of scales and associated symmetries, see text.
         The scenario where the breaking of LN and the generation of neutrino
         mass differences takes place at different scales, $\Lambda_{\rm LN}\gg \Lambda$,
         is not shown.}}
\end{center}
\end{figure}

The physical picture that we have in mind
is illustrated in Fig.~\ref{illustration}: Lepton number 
  is assumed to be broken at a very high scale (say, for instance, near 
  the GUT scale). 
  For hierarchical neutrino masses, we assume the large atmospheric neutrino mass 
  differences to be generated by a spurion VEV at the scale $\Lambda=\Lambda_{\rm LN}$
  (for almost degenerate neutrino masses, the situation $\Lambda\ll \Lambda_{\rm LN}$
   should be considered). The original lepton flavour symmetry $G_F \times U(1)_{\rm LN}$ 
  is thus broken to a subgroup $G_F'$ whose structure, as we will show, depends on
  the assumed neutrino mass pattern.
  The solar neutrino mass difference is related to the further breaking of $G_F'$ 
  which is assumed to happen at a lower scale $\Lambda'\ll \Lambda$. 
  (In this picture, the scale $\Lambda_{E}$ related to the generation of Yukawa couplings 
   for charged leptons always obeys $\Lambda_{E} \ll \Lambda$, but 
   $\Lambda_{E}>\Lambda'$, $\Lambda_{E} < \Lambda'$ or $\Lambda_E=\Lambda'$
   are possible.)
  At (or slightly above) the electroweak scale (below $\Lambda'$ and $\Lambda_{E}$), 
  the physics is described in terms of an ET sharing
  the gauge symmetry of the SM, with the flavour
  structure of higher-dimensional operators being dictated by the VEVs of spurion fields.

In the following, we are going to construct the non-linear representation
of lepton-flavour symmetry. We distinguish between different scenarios
for the neutrino mass hierarchy (``normal'', ``inverted'', ``degenerate'').
A few examples to illustrate the construction of ET operators 
below the scale $\Lambda'$ for radiative lepton-flavour transitions
and 4-lepton processes are discussed in section~\ref{sec3}. Finally, section~\ref{sec4} is 
devoted to systematic extensions beyond MLFV  (in the context of 
the non-linear representation of lepton flavour symmetry) along the lines proposed in \cite{Feldmann:2006jk}. 
We conclude with a brief summary in section~\ref{summary}.

\section{Non-linear representation}
Throughout this work, we assume that the possible right-handed neutrinos 
have masses of the order $\Lambda_{\rm LN}$ or higher, so that we can stick 
to  a scenario with minimal field content, where right-handed
neutrinos are assumed not to be part of the physical spectrum in the ET below the 
scale $\Lambda_{\rm LN}$.
In this case, the complete lepton flavour symmetry is described by,
 $$G_F \times U(1)_{\rm LN} \times U(1)_{\rm PQ} = 
 SU(3)_L\times SU(3)_{E_R} \times U(1)_L \times U(1)_{E_R} \,,$$
where the Peccei-Quinn symmetry $U(1)_{\rm PQ}$ distinguishes
right-handed up-quark fields $U_R$ from down-quark fields $D_R$
and charged leptons $E_R$ \cite{Peccei:1977ur}.
In the following discussion, we may ignore the $U(1)_{E_R}$ factor and 
concentrate on
\begin{align}
G_F \times U(1)_L = SU(3)_L\times SU(3)_{E_R} \times U(1)_L \,.
\end{align}
The breaking of this symmetry may be
described by two spurion fields\footnote{Here and in the following, unhatted
quantities denote scalar spurion fields with canonical mass dimension 1, whereas
hatted quantities denote dimensionless (Yukawa) couplings.}
\begin{eqnarray}
\hat Y_E  
&=&  V_{L}^\dagger \, {\rm diag} (y_e, y_\mu, y_\tau) \,, \\
\hat g_\nu \equiv \frac{g_\nu}{\Lambda_{\rm LN}}  
&=& \frac{\Lambda_{\rm LN}}{v^2} \, 
V_L^T \, U_{\rm PMNS}^* \, {\rm diag} (m_{\nu_1}, m_{\nu_2}, m_{\nu_2}) \, U_{\rm PMNS}^\dagger
\, V_{L}  \,,
\end{eqnarray}
where $V_{L} \in SU(3)_L \times U(1)_L$, with $V_{L}=1$ corresponding to the mass eigenbasis
for the charged  leptons, while $V_L=U_{\rm PMNS}$ (the PMNS
mixing matrix \cite{Pontecorvo:1967fh,Maki:1962mu}) 
defines the mass eigenbasis for neutrinos (right-handed transformations 
are not observable in the SM and set to unity). 
The matrix $\hat Y_E$ describes the SM Yukawa couplings of the charged leptons,
\begin{align}
 -{\cal L}_{\rm yuk} & = \bar L \, \hat Y_E \, H \, E_R + \mbox{h.c.}\,, 
\end{align}
and transforms as  $\hat Y_E \sim (3,\bar 3)_1$, where the numbers in
brackets refer to representations of $G_F$, and the index denotes the $U(1)_L$ charge associated
to left-handed lepton number. 
The matrix $g_\nu \sim (\bar 6,1)_{-2}$ appears in the higher-dimensional
operator,
\begin{equation} \label{Maj}
{\cal L}_{\rm Maj} = \frac{1}{2 \Lambda_{\rm LN}} \left( N^T \hat g_\nu \, N \right)
 \equiv \frac{1}{2\Lambda_{\rm LN}^2} \left( N^T g_\nu \, N \right)
+\mbox{h.c.}
\end{equation}
where $\epsilon_{\rm LN} = \Lambda/\Lambda_{\rm LN}$, and
\begin{equation}
N = \tilde H^\dagger  L   
\end{equation}
has vanishing quantum numbers under the complete SM gauge group.
Notice that the operator in (\ref{Maj}) is formally to be counted as dim-6 
when the coupling matrix $\hat g_\nu$
is promoted to a (scalar) spurion field $g_\nu$ with canonical mass dimension 1. 

If the scale $\Lambda_{\rm LN}$, associated with lepton-number violation,
is sufficiently large,  
$\Lambda_{\rm LN} \gg v$, 
the resulting neutrino masses
$m_{\rm Maj} \sim v^2/\Lambda_{\rm LN}$ are small, 
even if the spurion $\hat g_\nu$ has generic entries of order unity,
i.e.\ $\langle g_\nu \rangle = {\cal O}(\Lambda_{\rm LN})$.
Following the same strategy that led us to identify
the large top-Yukawa coupling in the quark sector, we may thus assume
that the large value $\langle g_\nu \rangle = {\cal O}(\Lambda_{\rm LN})$
is related to the largest eigenvalue in the neutrino mass matrix.
The remaining discussion depends on the assumed
hierarchy among the neutrino masses.
The experimental data on neutrino mixing (see e.g.\ \cite{Maltoni:2004ei}
and references therein),
with the two measured mass-squared differences $\Delta m^2_{\rm sol.}\ll
\Delta m^2_{\rm atm.}$,
allows for ``normal'' and ``inverted'' hierarchy
among the neutrino masses,
\begin{align}
 \mbox{normal: } & m_{\nu_1} , m_{\nu_2} \ll m_{\nu_3} \,, \\
 \mbox{inverted: } & m_{\nu_1} \sim m_{\nu_2} \gg m_{\nu_3} \,,
\end{align}
or even an almost degenerate case if the absolute mass scale for the
neutrinos is sufficiently large.
Notice that $\Delta m^2_{\rm sol}/\Delta m^2_{\rm atm} \sim 1/25 \sim \sin^2\theta_C \sim m_b/m_t$,
and thus the expansion parameter in the lepton sector is of similar size as in
the quark sector \cite{Feldmann:2008ja}.

\subsection{Normal neutrino-mass hierarchy}

Let us first discuss the case of normal hierarchy, for which
the leading structure of the flavour matrix $g$ in the
neutrino eigenbasis follows as,\footnote{The same result could be obtained 
in see-saw models, where the matrix $g_\nu$ may be constructed by integrating
out heavy right-handed neutrinos, interacting with left-handed neutrinos
and SM Higgs fields through Yukawa matrices $\hat Y_\nu$,
$$
  \frac{1}{\Lambda_{\rm LN}} \, \hat g_\nu 
  \propto (\hat Y_\nu)^* \, (M_R)^{-1} \, \hat Y_\nu^\dagger \,.
$$
Assuming that, analogously to the Yukawa matrix in the up-quark sector
\cite{Feldmann:2008ja}, 
the neutrino Yukawa matrix has one large entry, that without loss of generality
may be chosen in the lower right corner, one recovers (\ref{gnormal}). }
\begin{align}
 \mbox{normal: } \quad & g_\nu \simeq
\left(
 \begin{array}{ccc}
   0 & 0 & 0 \\
   0 & 0 & 0 \\
   0 & 0 & g
 \end{array}
\right) \Lambda_{\rm LN} \,,
\label{gnormal}
\end{align}
with 
\begin{align}
 g = \frac{\Lambda_{\rm LN}}{v^2} \, m_{\nu_3}  \simeq \frac{\Lambda_{\rm LN}}{v^2} \,
  \sqrt{\Delta m^2_{\rm atm}} = {\cal O}(1) \,.
\end{align}
The matrix (\ref{gnormal}) breaks the original flavour symmetry as
\begin{align}
 G_F \times U(1)_{\rm L} \to 
 G_F' = SU(2)_{L} \times SU(3)_{E_R} \times U(1)_{\rm L^{(2)}} \times Z_2\,.
\label{break-normal}
\end{align}
Here, the combination
\begin{align}
 L^{(2)} = \frac23 \, L + \frac{2}{\sqrt3} \, T_L^8 
= \left( \begin{array}{ccc} 1 & 0&0 \\ 0 & 1 & 0 \\ 0 & 0 & 0
                  \end{array} \right) 
\end{align}
is the generator for (left-handed) lepton number in the 2-generation sub-space,
and the discrete $Z_2$ symmetry is represented by a particular group 
element of $U(3)_L $,  
\begin{align}
V_1 & =  \left( \begin{array}{ccc} 1 & 0&0 \\ 0 & 1 & 0 \\ 0 & 0 & e^{ i \pi }
                  \end{array} \right) 
\nonumber 
\,, \quad 
\end{align}
which commutes with $SU(2)_L\times U(1)_{L^{(2)}}$ transformations
and leaves the VEV for $g_\nu$ in (\ref{gnormal})
invariant.

The Goldstone modes $\Pi_L^a$ ($a=4\ldots 8$) associated to the 5 broken generators
of the continuous $SU(3)_L$ symmetry
define the non-linear representation of the spurion $g_\nu$,
\begin{align}
 \hat g_\nu = {\cal U}^*(\Pi_L) \, \left( \begin{array}{cc} 
g_{\nu}^{(2)}/\Lambda_{\rm LN} & \begin{array}{c} 0 \\ 0 \end{array} \\
\begin{array}{cc} 0 & 0 \end{array} & g
\end{array} \right) {\cal U}^\dagger(\Pi_L) \,.
\label{Ucal}
\end{align}
They are introduced in the standard parameterization~\cite{Coleman:1969sm}, 
\begin{equation}
{\cal U}(\Pi_L) = \exp \left(\frac{i}{\Lambda_{\rm LN}} \, \sum_{a=4}^8 \,
 T^a \, \Pi_L^a \right) \,,
\label{UL}
\end{equation}
and transform under the full flavour symmetry group $G_F$ in
a non-linear way \cite{Feldmann:2008ja}.
The remaining spurion $g_{\nu}^{(2)}$ has canonical mass dimension,
carries the charge $L^{(2)}=-2$, and transforms trivially
under $Z_2$. It is represented by a 
complex symmetric $2\times 2$ matrix $g_\nu^{(2)}$
which can be decomposed as
\begin{align}
 g_\nu^{(2)} &= i \sigma_2 \left( \phi_1 + i \, \phi_2 \right)
= i \sigma_2 \, \sum_{a=1}^3 \left( \phi_1^a + i \, \phi_2^a \right) \sigma_a \,,
\end{align}
where $\phi_{1,2}$ are traceless hermitian matrices 
transforming both as triplets under $SU(2)_L$. 
At the scale $\Lambda^\prime$ the field  $g_\nu^{(2)} $ acquires a VEV, 
and the eigenvalues of this VEV determine the
two small neutrino masses $m_{\nu_{1,2}}$, defining $\Delta m^2_{\rm sol.}$
in the normal-hierarchy scenario.
The 5 Goldstone-modes, the large eigenvalue $g$, 
and the six real parameters in $g_{\nu}^{(2)}$ add up to 12 degrees of freedom
describing the complex symmetric matrix $g_\nu$.
Following \cite{Feldmann:2008ja}, we introduce the projections 
$\Xi_L$ and ${\cal U}_L^{(2)}$ via
\begin{equation}
 {\cal U}(\Pi_L)_{ij} = (\Xi_L)_i \, \delta_{j3}
+ \sum_{k=1,2} ({\cal U}_L^{(2)})_{ik} \, \delta_{kj} \,. 
\end{equation}
The neutrino-mass operator (\ref{Maj})
in the non-linear representation can then be written as
\begin{equation} \label{Maj:normal}
{\cal L}_{\rm Maj} =  
 \frac{g}{2\Lambda_{\rm LN}} \left( N^T \Xi_L^* \, \Xi_L^\dagger N \right)
+
\frac{1}{2\Lambda_{\rm LN}^2} 
\left( N^T {\cal U}_L^{(2)*} \, g_\nu^{(2)} \, {\cal U}_L^{(2)\dagger} N \right)
+ \mbox{h.c.} \,,
\end{equation}
where the neutrino mass hierarchy is now manifest, with
$\Delta m^2_{\rm atm}$ arising from a dim-5 term, and $\Delta m^2_{\rm sol}$
from a dim-6 operator.

Analogously, the Yukawa matrix for the charged leptons may be decomposed into 
\begin{align}
  \hat Y_E & \equiv \frac{1}{\Lambda_{\rm LN}} \, {\cal U}(\Pi_L) \left( \begin{array}{c}
                                            Y_E^{(2)}\\ \xi_{E_R}^\dagger
                                           \end{array}
 \right)
\equiv \frac{1}{\Lambda_{\rm LN}} \left(
 {\cal U}_L^{(2)} \, Y_E^{(2)} + \Xi_L \, \xi_{E_R}^\dagger \right)
\,,
\label{YErepr}
\end{align}
where we introduced the $G_F'$--irreducible spurions
$Y_E^{(2)} \sim (2,\bar 3)_{1,0}$ and $\xi_{E_R}^\dagger \sim (1,\bar 3)_{0,1}$, 
with the first index refering to the $U(1)_{L^{(2)}}$ charge,  and the second index 
to the $Z_2$ representation ($0=$ trivial, $1=$ fundamental).
For every charged lepton, the Yukawa terms are thus already dim-5,
\begin{align}
 -{\cal L}_{\rm yuk} & = 
  \frac{1}{\Lambda_{\rm LN}} \, \bar L \, {\cal U}_L^{(2)} \, Y_E^{(2)} \, H \,  E_R
 +  \frac{1}{\Lambda_{\rm LN}} \, \bar L\,\Xi_L \, \xi_{E_R}^\dagger \, H \, E_R
 + \mbox{h.c.} \,,
\label{Lyuk2}
\end{align}
and
the PMNS matrix is identified as 
\begin{align}
 U_{\rm PMNS}^\dagger & = 
 \left( \begin{array}{c c}
    V_{\nu_L}^{(2)} & \begin{array}{c} 0 \\ 0 \end{array} \\ 
    \begin{array}{cc} 0 & 0 \end{array} & 1 \end{array} \right)
{\cal U}^\dagger(\Pi_L) \, V_L^\dagger 
= \left( 
\begin{array}{c}
V_{\nu_L}^{(2)} \, {\cal U}_L^{(2)\dagger} \\
\Xi_L^\dagger
\end{array} \right) \, V_L^\dagger
\,,
\label{PMNS1}
\end{align}
where $V_{\nu_L}^{(2)} \in U(2)_L $ diagonalizes $g_\nu^{(2)}$.

\paragraph{Mass eigenbasis for charged leptons}

Often, the structure of the neutrino mass matrix is considered
in the mass eigenbasis for the charged leptons.
Let us approximate the PMNS matrix by the
so-called tri-bimaximal mixing form \cite{Harrison:2002er},
\begin{align}
 U_{\rm PMNS} & {} \simeq 
R_{23}\left(-\frac{\pi}{4}\right) \, R_{12}\left(\arcsin \frac{1}{\sqrt3}\right) 
= \left(
\begin{array}{ccc}
  \sqrt{2/3} & \sqrt{1/3} & 0 \\
  -\sqrt{1/6} & \sqrt{1/3} & -\sqrt{1/2} \\
  -\sqrt{1/6} & \sqrt{1/3} &  \sqrt{1/2}
\end{array}
\right) 
\label{U0}
\end{align}
and ignore Dirac and Majorana phases.
In the limit $\Lambda_{\rm LN}\to \infty$, the leading term 
for the neutrino mass matrix (\ref{gnormal}) 
in the charged-lepton eigenbasis
then reads
\begin{align}
 \langle \hat g_\nu \rangle \to 
U_{\rm PMNS} \, \langle \hat g_\nu \rangle \, U_{\rm PMNS}^T
 \simeq
 \frac{g}{2} \left( \begin{array}{ccc}
 0 & 0 & 0 \\ 0 & 1 & -1 \\ 0 & -1 & 1
\end{array}\right) \,.
\end{align}
In this basis the neutrino matrix exhibits an apparent $U(1)$ symmetry,
where lepton number in the electron sector ($L_e$) is still (approximately) conserved
(see, for instance, the discussion in \cite{Choubey:2004hn} and references
therein). In our framework, the $L_e$ symmetry is realized by a particular
linear combination of $L^{(2)}$, $T_L^1$ and $T_L^3$ (and $E_R$). We should
stress at this point, that our approach of identifying the residual
flavour symmetry $G_F'$ in the limit $\Lambda_{\rm LN} \to \infty$
is basis independent and more general than finding approximately
conserved lepton-flavour charges in the charged-lepton eigenbasis.


\subsection{Inverted neutrino-mass hierarchy}

Similarly, in the
case of inverted hierarchy,
the leading structure of the flavour matrix $g_\nu$ in the
neutrino eigenbasis reads,
\begin{align}
 \mbox{inverted: } \quad  & g_\nu \simeq
\left(
 \begin{array}{ccc}
   g & 0 & 0 \\
   0 & g & 0 \\
   0 & 0 & 0
 \end{array}
\right) \Lambda_{\rm LN} \,,
\label{ginvert}
\end{align}
with 
\begin{align}
 g = \frac{\Lambda_{\rm LN}}{v^2} \, m_{\nu_{1,2}} \simeq \frac{\Lambda_{\rm LN}}{v^2} \, 
 \sqrt{\Delta m^2_{\rm atm.}} = {\cal O}(1) \,.
\end{align}
It breaks the original flavour symmetry,
\begin{align}
 G_F \times U(1)_L \to G_F' =  SO(2)_L \times SU(3)_{E_R} \times U(1)_{L_3} \,.
\label{break-invert}
\end{align}
Here,  the unbroken $SO(2)_L$ generator is 
given by $T^2$ from $SU(3)_L$, and 
the $U(1)_{L_3}$ transformations are generated by the linear combination
\begin{align}
 L_3 = \frac13 \, L - \frac{2}{\sqrt3} \, T_L^8 \,.
\end{align}
The remaining 7 Goldstone modes are then introduced by
the exponential
$$
  {\cal U}(\Pi_L) = \exp \left(\frac{i}{\Lambda_{\rm LN}} \, \sum_{a\neq 2} \,
 T^a \, \Pi_L^a \right) \,,
$$
and the representation of the flavour matrix $g_\nu$ reads
\begin{align}
 \hat g_\nu & = {\cal U}^*(\Pi_L) \, 
\left( 
 \begin{array}{cc} 
  g \, {\boldsymbol{1}} + \tilde g_\nu^{(2)}/\Lambda_{\rm LN}  & 
  \begin{array}{c} 0 \\ 0 \end{array}  \\
   \begin{array}{cc} 0 & 0 \end{array} & \phi_3/\Lambda_{\rm LN}
 \end{array} 
\right)
 {\cal U}^\dagger(\Pi_L) \,.
\end{align}
Here the new spurion $\tilde g_\nu^{(2)}$ is 
a real symmetric traceless $2\times 2$ matrix transforming as $(2,1)_{0}$ under $G_F'$. 
At the scale $\Lambda^\prime$ $\tilde g_\nu^{(2)}$  acquires a VEV, 
whose eigenvalue determines the mass-splitting between $m_{\nu_1}$ and $m_{\nu_2}$
giving rise to $\Delta m^2_{\rm sol.}$.
The complex spurion $\phi_3$ is a singlet under $SO(2)_L $
with $L_3=-2$, and
its absolute value determines the small neutrino mass $m_{\nu_3}$.
The large eigenvalue $g$, the 7 Goldstone modes and the
four real parameters for $\tilde g_\nu^{(2)}$, $\phi_3$ 
add up to 12 parameters necessary to describe the complex 
symmetric $3\times 3$ matrix $g_\nu$.
The remaining discussion is completely analogous to the case
of normal neutrino-mass hierarchy with the appropriate changes
from $U(2)_L \times Z_2$ to 
     $SO(2)_L \times U(1)_{L_3} $ transformations.
In particular, the residual spurions for the charged-lepton Yukawa matrix
now transform as 
$Y_E^{(2)} \sim (2,\bar 3)_{0}$, and  
$\xi_R^\dagger \sim (1,\bar 3)_{1}$.

\subsection{Almost degenerate neutrino masses}

Degenerate neutrino masses are obtained from
\begin{align}
 \mbox{degenerate: } \quad  & g_\nu \simeq
\left(
 \begin{array}{ccc}
   g & 0 & 0 \\
   0 & g & 0 \\
   0 & 0 & g
 \end{array}
\right) \Lambda_{\rm LN} \,,
\label{gdegen}
\qquad
 g = \frac{\Lambda_{\rm LN}}{v^2} \, \bar m_{\nu} = {\cal O}(1) \,.
\end{align}
This breaks the original flavour symmetry,
\begin{align}
 G_F \times U(1)_L \to G_F' =  SO(3)_L \times SU(3)_{E_R} \,.
\end{align}
The remaining 6 Goldstone modes are introduced by
the exponential
$$
  {\cal U}(\Pi_L) = \exp \left(\frac{i}{\Lambda_{\rm LN}} \, \sum_{a\neq 2,5,7} \,
 T^a \, \Pi_L^a \right) \,,
$$
and the representation of the flavour matrix $g_\nu$ reads
\begin{align}
 \hat g_\nu & = {\cal U}^*(\Pi_L) 
\left( g \, {\boldsymbol{1}} + \frac{1}{\Lambda_{\rm LN}} \,
  \check g_\nu \right)
 {\cal U}^\dagger(\Pi_L) \,.
\end{align}
Here the new spurion $\check g_\nu$ is represented by 
a real symmetric traceless $3\times 3$ matrix transforming 
as $(5,1)$ under $G_F'$,
whose eigenvalue determine the neutrino mass-splittings. 
The residual spurion for the charged-lepton Yukawa matrix
transforms as $\check Y_E \sim (3,\bar 3)$.

Now, the largest neutrino mass difference $\Delta m^2_{\rm atm}$ has to be assigned to a
VEV for the spurion $\check g_\nu$ at a scale $\Lambda \ll \Lambda_{\rm LN}$,
such that
$$
  \Delta m^2_{\rm atm}/\bar m_\nu^2 = {\cal O}(\Lambda/\Lambda_{\rm LN}) \,,
$$
whereas $\Delta m^2_{\rm sol}$ would be generated at even smaller scales,
$\Lambda' \ll \Lambda$.
Taking
\begin{align}
 & \check g_\nu \simeq
\left(
 \begin{array}{ccc}
   -\check g & 0 & 0 \\
   0 & -\check g & 0 \\
   0 & 0 & 2 \check g
 \end{array}
\right) \Lambda \,,
\label{gcheck}
\end{align}
the flavour symmetry is further broken,\footnote{Notice that the same symmetry
breaking in the left-handed sector could be obtained from a VEV for the charged lepton Yukawa
spurion $\langle \check Y_E\rangle \to {\rm diag}(0,0,y_\tau \, \Lambda)$
(which also breaks $SU(3)_R$).
One could even speculate that in the scenario 
with degenerate neutrino spectrum,
the generation of the $\tau$ Yukawa coupling and
the atmospheric neutrino mass difference are related such that 
$\Delta m^2_{\rm atm}/\bar m_\nu^2 \sim m_\tau/v_{\rm SM}$ 
implying $\bar m_\nu \sim {\cal O}(0.5$~eV$)$, which happens to
be close to the present upper experimental bound.
}
\begin{align}
 G_F' \to  G_F'' = SO(2)_L \times SU(3)_{E_R} \times Z_2 \,.
\end{align}
Two new Goldstone modes are introduced by
the exponential
$$
  {\cal U}(\check \Pi_L) = \exp \left(\frac{i}{\Lambda} \, \sum_{a=5,7} \,
 T^a \, \check \Pi_L^a \right) \,,
$$
and the representation of the flavour matrix $\check g_\nu$ reads
\begin{align}
 \frac{\check g_\nu}{\Lambda} & = {\cal U}^*(\check \Pi_L) 
\left( 
 \begin{array}{cc} 
  - \check g \, {\boldsymbol{1}} + \check g_\nu^{(2)}/\Lambda  & 
  \begin{array}{c} 0 \\ 0 \end{array}  \\
   \begin{array}{cc} 0 & 0 \end{array} & 2 \check g
 \end{array} 
\right)
 {\cal U}^\dagger(\check \Pi_L)  \,.
\end{align}
Here the eigenvalues of 
the new spurion $\check g_\nu^{(2)} \sim (2,1)_0$ 
determine $\Delta m^2_{\rm sol}$.
The residual spurions for the charged-lepton Yukawa matrix
transform as $\check Y_E^{(2)} \sim (2,\bar 3)_0$
and $\check \xi_{E_R}^\dagger \sim (1,\bar 3)_1$ under $G_F''$.

\section{Effective theory at $\Lambda'$ and MLFV}

\label{sec3}

In the following, we discuss a few examples of how to construct MLFV operators in
the ET, starting with the non-linear representation
of spurion fields.\footnote{
We remind the reader that we stick to the case of minimal (SM) field content, here. 
For a more general discussion of
MLFV, see also \cite{Davidson:2006bd}.}
We pay particular 
attention on how to obtain the effective operators at or slightly below
the intermediate scales, $\Lambda',\Lambda_E, \ldots$,
in terms of the spurion fields 
which have been introduced close to the high-energy scale,
 $\Lambda=\Lambda_{\rm LN}$.

\subsection{Example: radiative decays $\ell \to \ell' \gamma$ }

The discussion of radiative LFV decays ($\tau \to \mu(e)\gamma$,
$\mu \to e\gamma$) is very similar to the analogous quark decays, see
\cite{Feldmann:2008ja}. Let us concentrate on the case of normal neutrino
hierarchy, first, and assume for simplicity that we only have one
intermediate scale $\Lambda' \sim \Lambda_E$, where the residual
spurions of $G_F'$ acquire their VEVs.
A typical MLFV operator in the effective Lagrangian \emph{above} 
the scale $\Lambda'$ would read
\begin{equation}
 {\cal O}_{\rm eff} = \frac{1}{\Lambda_{\rm LN}^3} \,
 (\bar L \,  \Xi_{L} \, H \, \sigma_{\mu\nu} \
  \xi_{E_R}^\dagger  E_R) \, F^{\mu\nu} + {\rm h.c.}
\label{Oeffhigh}
\end{equation}
where $F^{\mu\nu}$ is the field strength tensor for 
the gauge field $B_\mu$ associated to hypercharge in the SM 
(a similar term with the $SU(2)_L$ field strength
 $W^3_{\mu\nu}$ is also present).
It contributes at tree-level to $\ell \to \ell'\gamma$, when
$\xi_{E_R} \to \langle \xi_{E_R} \rangle \sim \Lambda'$ 
and $H \to \langle H \rangle =v$, however,
with a very small pre-factor of order 
$(v \, \Lambda')/\Lambda_{\rm LN}^3$.

\begin{figure}[t!pt]
 \begin{center}
  \includegraphics[width=0.28\textwidth]{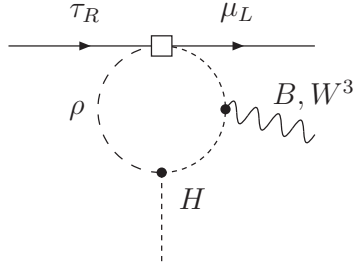}
\vspace{0.8em}
\parbox{0.9\textwidth}{\small
\caption{\label{fig:1loop} \small Example for 
1-loop matching contribution to ${\cal O}_{\rm eff}$ in (\ref{Oeff}). Here $\rho$ is the heavy scalar degree of freedom 
 appearing in the spurion $\xi_{E_R}$, which interacts with
 the SM Higgs and charged leptons 
 through the scalar potential and the dimension-5 Yukawa terms (\ref{Lyuk2}).}
}
\end{center}
\end{figure}

On the other hand, below the scale $\Lambda'$, the heavy scalar
degrees of freedom in the spurion field $\xi_{E_R}$ have to
be integrated out. Taking into account scalar couplings\footnote{
%
%
Focusing on the relevant part of the scalar potential which involves
$\xi_{E_R}$ and the SM Higgs, we may write
$$
V_\xi = - \mu_\xi^2 \, \xi_{E_R}^\dagger\xi_{E_R}
        + \frac{\lambda_\xi}{2} \, (\xi_{E_R}^\dagger\xi_{E_R})^2
        + \kappa \, (\xi_{E_R}^\dagger\xi_{E_R}) \, (H^\dagger H)
$$
which is minimized for $\mu_\xi^2 = \kappa \, v^2 + \lambda_\xi \, \langle \xi_{E_R} \rangle$.
Parameterizing the remaining heavy degree of freedom as $\xi_{E_R}^\dagger - \langle \xi_{E_R}^\dagger \rangle= (0,0,\rho)$, we obtain the effective potential,
$$
V_\xi = \kappa \, (\rho + \langle \xi_{E_R}^\dagger\rangle )^2 \, (H^\dagger H)
 + \frac12 \, \rho \, (\rho + 2 \langle \xi_{E_R}^\dagger\rangle ) \,
        \left( \lambda_\xi \, \rho \, (\rho + 2 \langle \xi_{E_R}^\dagger\rangle)-2 \kappa v^2
  \right)
$$
which, among others, contains the coupling $\propto  \rho H^\dagger H$
used in Fig.~\ref{fig:1loop}.
%
%
} 
between
$\xi_{E_R}$ and $H$ together with the dim-5 Yukawa term
involving $\xi_{E_R}$ in (\ref{Lyuk2}), one can generate loop diagrams as
shown in Fig.~\ref{fig:1loop}, which \emph{below} the scale
$\Lambda'$ induce effective operators of the form
\begin{equation}
 {\cal O}_{\rm eff} = \frac{1}{(4\pi\Lambda')^2} \,
 (\bar L \,  \Xi_{L} \, H \, \sigma_{\mu\nu} \
  \frac{\langle \xi_{E_R}^\dagger\rangle}{\Lambda_{\rm LN}} \, E_R) \, F^{\mu\nu} + {\rm h.c.}
\label{Oeff}
\end{equation}
which have the same flavour structure as (\ref{Oeffhigh}), but
a somewhat larger pre-factor of order $v/(4\pi\Lambda')^2 \cdot \Lambda'/\Lambda_{\rm LN}$,
where $\Lambda'/\Lambda_{\rm LN} \sim m_\tau/v$.

After changing to the mass eigenbasis for the charged leptons, 
using (\ref{YErepr},\,\ref{PMNS1}),
the effective operator (\ref{Oeff}) exhibits the flavour structure
\begin{equation}
\left( V_L \, \Xi_{L} \, \xi_{E_R}^\dagger \right)_{ij}
=  U_{i3} U_{j3}^* \, (y_\ell)_j \, \Lambda_E \,,
\label{s1}
\end{equation}
which reproduces the leading term for the result discussed in \cite{Cirigliano:2005ck}, in
the limit $\Delta m^2_{\rm sol.} \ll \Delta m^2_{\rm atm.}$. The expression
can be traced back to the effective quantity
\begin{align}
\hat \Delta = V_L \, \hat g_\nu^\dagger \hat g_\nu  V_L^\dagger 
             - 1/3 \, {\rm tr}(\hat g_\nu^\dagger \hat g)
= \frac{\Lambda_{\rm LN}^2}{v^4} 
 \left( U_{\rm PMNS} \, {\rm diag}[m_\nu^2] \, U_{\rm PMNS} ^\dagger
  - \frac13 \, {\rm tr}[m_\nu^2] \right) \,, 
\label{Delta}
\end{align}
which in the limit $m_{\nu_3}\gg m_{\nu_{1,2}}$ is given by, cf.~(\ref{gnormal}),
\begin{align}
  \hat \Delta_{ij} 
\ \stackrel{m_{\nu_3}\gg m_{\nu_{1,2}}}{\longrightarrow} \
g^2 \left( U_{i3} U_{j3}^* - \frac{\delta_{ij}}{3} \right)
\ \simeq  \ \frac{g^2}{6}  \left( 
\begin{array}{ccc}
  -2& 0 & 0 \\ 0 & 1 & -3 \\ 0 & -3 & 1
\end{array}
\right)_{ij}\,,
\label{D1}
\end{align}
where in the last line we inserted the approximation of
the PMNS matrix for tri-bimaximal mixing (\ref{U0}).
Sub-leading effects are induced by  $g_\nu^{(2)}$, which can be seen
by either including the corresponding $1/\Lambda$ corrections in (\ref{Delta}),
or by directly inserting additional powers of $g_\nu^{(2)}$ in 
effective operators like (\ref{Oeff}) as allowed by flavour symmetry.
The discussion for the inverted hierarchy is completely analogous with,
cf.~(\ref{ginvert}),
\begin{align}
 \hat \Delta_{ij} 
\
 \stackrel{m_{\nu_3}\ll m_{\nu_{1,2}}}{\longrightarrow} \
g^2 \left( \frac{\delta_{ij}}{3} - U_{i3} U_{j3}^*\right)
\ \simeq \ \frac{g^2}{6}  \left( 
\begin{array}{ccc}
  2& 0 & 0 \\ 0 & -1 & 3 \\ 0 & 3 & -1
\end{array}
\right)_{ij}
\,.
\end{align}
Finally, for degenerate neutrino masses we obtain
$\hat \Delta \to 0$.

\subsection{Example: 4-lepton processes}

Flavour-violating 4-lepton processes are interesting, because different chirality
structures can be experimentally constrained by a Dalitz-plot analysis and/or
angular distributions \cite{Dassinger:2007ru,Matsuzaki:2007hh,Giffels:2008ar}, 
and this information may be used to distinguish between different NP models.

In the linear version of MLFV, see \cite{Cirigliano:2005ck,Dassinger:2007ru}, 
besides the effective quantity $\hat\Delta_{ij}$ discussed in the previous
subsection, additional flavour structures arise,
which can be expressed in terms of the tensor
$G_{ij}^{kl}$, describing the 27-plet in the reduction of $(\bar 6,1)\times (6,1)
= 1 + 8 + 27$,
\begin{eqnarray}
 \hat G_{ij}^{kl} &=& (V_L^* \hat g_\nu V_L^\dagger)_{ij} \, (V_L \hat g_\nu^* V_L^T)^{kl} 
  - \frac{1}{12} \left(\delta_i^k \delta_j^l + \delta_i^l \delta_j^k \right) 
     {\rm tr}(\hat g_\nu^\dagger \hat g_\nu) \nonumber\\[0.2em]
&& {} 
  - \frac{1}{5} \left( 
      \delta_i^a \delta_b^l \delta_j^k + \delta_j^a \delta_b^l \delta_i^k
    + \delta_i^a \delta_b^k \delta_j^l + \delta_j^a \delta_b^k \delta_i^l
   \right) \, \hat \Delta^b{}_a \,.
\end{eqnarray}
Here $\hat G_{ij}^{kl}=  \hat G_{ji}^{kl}= \hat G_{ij}^{lk}$, and
$
  \sum_i \, \hat G_{ij}^{il} = 0
$. 
The 27-plet appears in purely left-handed operators as
$$
 G_{ij}^{kl} \, (\bar L_k \, L^i)(\bar L_l \, L^j) \,.
$$

In the non-linear version of MLFV, the discussion is somewhat different.
The elementary flavour-symmetry invariant building blocks for the leading
left-handed operators are
\begin{align}
&  (\bar L \, L) \,, \qquad (\bar L \, \Xi_L) \,, \qquad (\Xi_L^\dagger L) \,.
\end{align}
After changing to the mass eigenbasis, the first term remains flavour diagonal,
whereas
\begin{align}
&  (\bar L \, \Xi_L) \to  (\bar L \, V_L \, \Xi_L) =   \bar L_i \, U_{i3} \,, \cr
& (\Xi_L^\dagger\bar L) \to (\Xi_L^\dagger \, V_L^\dagger \, \Xi_L) =  U^*_{j3} \, L_j \,,
\end{align}
generate the same flavour factors as in $\hat \Delta_{ij}$ in (\ref{s1}).
At tree level, the leading flavour coefficients in purely left-handed 4-lepton operators
are thus determined by structures like
\begin{align}
 &  \frac{1}{\Lambda_{\rm LN}^2} \, (\bar L \, \gamma_\mu \, L) \, 
                                   (\bar L \, \Xi_L) \, \gamma^\mu \,(\Xi_L^\dagger L) \,,
\cr 
 &  \frac{1}{\Lambda_{\rm LN}^2} \,
  (\bar L \, \Xi_L) \,\gamma_\mu \, (\Xi_L^\dagger L) \, \gamma^\mu \, (\bar L \, \Xi_L) \,(\Xi_L^\dagger L) \,.
\end{align}
Loop diagrams, contributing to flavour-violating 4-lepton processes and
involving the heavy degrees of freedom associated with the breaking of 
$G_F'$ at the scale $\Lambda'$, 
require at least two insertions of $1/\Lambda_{\rm LN}$ suppressed operators.
For instance, in the case of normal hierarchy,
we obtain terms like 
\begin{align}
 & \frac{1}{(4\pi \Lambda')^2} \,
(\bar L \, L) \, (\bar L \, {\cal U}_L^{(2)}  
\frac{\langle g_\nu^{(2)\dagger}g_\nu^{(2)}\rangle}
{\Lambda_{\rm LN}^2} \, {\cal U}_L^{(2)\dagger} L)  \,, 
\cr 
& \frac{1}{(4\pi \Lambda')^2} \, (\bar L \, L) \, (\bar L \,{\cal U}_L^{(2)} 
\frac{\langle Y_E^{(2)} Y_E^{(2)\dagger}\rangle}
{\Lambda_{\rm LN}^2}\, {\cal U}_L^{(2)\dagger} L) \,,
 \quad \mbox{etc.}
\end{align}

Including right-handed fields, the leading tree-level flavour structures 
are obtained from 4-lepton operators of the form
\begin{align}
 &  \frac{1}{\Lambda_{\rm LN}^2} \, (\bar E_R \, E_R) \, (\bar L \, \Xi_L) \,(\Xi_L^\dagger L) \,.
\end{align}
Again, insertions of sub-leading operators in loop diagrams with
heavy spurion degrees of freedom 
lead to additional structures, like
\begin{align}
& \frac{1}{(4\pi \Lambda')^2} \,
  (\bar E_R \, \frac{\langle \xi_{E_R} \xi_{E_R}^\dagger \rangle}
 {\Lambda_{\rm LN}^2} \, E_R) \, (\bar L \,  L)
\quad \mbox{etc.,}
\end{align}
and similarly for the inverted and degenerate case.

\section{Beyond MLFV}

\label{sec4}

A systematic procedure to include deviations from the MFV assumption 
within the ET framework (next-to-minimal flavour violation, nMFV)
has been proposed in \cite{Feldmann:2006jk}
(for alternative approaches, see \cite{Agashe:2005hk,Fitzpatrick:2007sa,Davidson:2007si}).
However, the formalism has been worked out for quark decays in the linear formulation of MFV, only. 
In the following, we are going to apply the nMFV ansatz to the lepton sector within the non-linear formulation of lepton-flavour violation (nMLFV).

\subsection{Normal hierarchy}

\begin{table}[t!pbh]
\begin{center}
 \small
 \begin{tabular}{| l || c | c | c | c |}
\hline
   & $E_R$ & $({\cal U}_L^{(2)\dagger} L)$ &  $(\Xi_L^\dagger L)$ & $({\cal U}_L^{(2)\dagger} L)^*$  \\
\hline
$\bar E_R$ & $Z_E \sim(1,8)_{0,0}$ 
           & $Y_E^{(2)\dagger} \sim (2,3)_{-1,0}$
           & $\xi_{E_R} \sim (1,3)_{0,1}$ 
           & $X_E^{(2)\dagger} \sim (2,3)_{1,0}$ 
\\
$(\bar L \, {\cal U}_L^{(2)})$ 
           & $\bullet$ 
           & $Z_L^{(2)} \sim (3,1)_{0,0}$
           & $\chi_L \sim (2,1)_{1,1}$ 
           & $g_\nu^{(2)^*} \sim (3,1)_{2,0}$ 
\\
$(\bar L \, \Xi_L)$ 
           & $\bullet$ 
           & $\bullet$ 
           &  $ \sim (1,1)_{0,0}$ 
           &  $\chi_L \sim (2,1)_{1,1}$ 
\\
$(\bar L \, {\cal U}_L^{(2)})^*$ 
           &  $\bullet$ 
           &   $\bullet$ 
           &  $\bullet$
            & $Z_L^{(2)} \sim (3,1)_{0,0}$
\\
 \hline
 \end{tabular}
\caption{\label{tab:nMFVnormal} 
Possible bi-linear combinations of fundamental fermion fields and
the associated spurion fields (normal hierarchy,
$G_F'=SU(2)_L \times SU(3)_{E_R} \times U(1)_{L^{(2)}} \times Z_2$).
}
\end{center}
\end{table}

The basic idea of nMLFV is to introduce additional spurion fields that can
couple to fundamental fermion bi-linears appearing in higher-dimensional gauge-invariant operators.
Let us discuss the case of normal neutrino mass hierarchy, first.
The basic fermion fields with definite transformations under $G_F'$ are
\begin{align}
 & E_R \sim (1,3)_{0,0} \,,  & & ({\cal U}_L^{(2)\dagger} L)   \sim (2,1)_{+1,0} \,, \cr
 & (\Xi_L^\dagger L) \sim (1,1)_{0,1} \,, & & 
  ({\cal U}_L^{(2)\dagger} L)^* \sim (2,1)_{-1,0} \,.
\end{align}
Out of these four fields, we can construct all possible
bi-linear flavour structures, as shown in Table~\ref{tab:nMFVnormal}.
In nMLFV each of these combinations corresponds to an independent spurion
field. Besides the spurions
$g_\nu^{(2)}$, $\xi_{E_R}^\dagger$ and $Y_E^{(2)}$, appearing in the
non-linear formulation of MLFV, we also obtain
new spurion fields $\chi_L$, $X_E^{(2)}$, $Z_L^{(2)}$ and $Z_E$. 
In nMLFV, we may thus consider new operators like, for instance,
\begin{align}
&
 \frac{1}{\Lambda^2}
 \, (\bar L \, {\cal U}_L^{(2)} \, \chi_L \,\gamma^\mu \, \Xi_L^\dagger L)
  \, (H^\dagger D_\mu H) \,,
\quad 
 \frac{1}{\Lambda^2}
 \, (\bar L \, {\cal U}_L^{(2)} \, \sigma^{\mu\nu} \, H \, X_E^{(2)} E_R)
  \, B_{\mu\nu} \,, \quad \mbox{etc.}
\end{align}
Insertions of the spurion $\chi_L$ induce significant 
contributions to $\tau \to e\gamma$, 
where the leading MLFV contributions vanishes, because
$U_{13}\simeq 0$, see (\ref{s1},\ref{D1}). The spurions $Z_E$ and
$X_E$ induce new lepton flavour-violating structures 
involving right-handed leptons.
The set of new spurions can thus be used to parameterize deviations from
the correlations between different LFV observables as one would
predict in MLFV \cite{Cirigliano:2006su,Branco:2006hz}.

As explained in \cite{Feldmann:2006jk}, the new spurion fields can
also appear in operators whose gauge structure is already present
in MLFV, for instance
\begin{align}
  \frac{1}{\Lambda^3} \, \bar N^T {\cal U}_L^{(2)*}
    (\chi_L^* \, \chi_L^\dagger)
  \ {\cal U}_L^{(2)\dagger} N \,.
\end{align}
A minimal constraint on the new spurion fields then 
follows from self-consist\-ency relations for
those combinations of old and new spurion fields 
that transform as the original MLFV spurions. 
As a consequence, the power-counting for the new spurion fields is 
limited from above by the phenomenology of lepton masses and mixing. 
In this context, an advantage of the non-linear formulation of MFV is that products of spurion fields in the effective Lagrangian are always suppressed by higher powers of $1/\Lambda$ compared to single spurion insertions. 
Therefore, we can safely restrict the discussion to products of two spurion fields. 
In the case of normal hierarchy this yields
the following set of inequalities: 
\begin{align}
 \left( \frac{g_\nu^2}{\Lambda}\right)_{ij} 
& \gtrsim \
 \left\{
\left( \frac{\chi_L^* \, \chi_L^\dagger}{\Lambda^2}\right)_{ij} \,,
\left( \frac{X_E^{(2)} \, Y_E^{(2)\dagger}}{\Lambda^2}\right)_{ij} 
\right\} \,,
\nonumber \\[0.3em]
\left( \frac{\xi_{E_R}^\dagger}{\Lambda}\right)_{ij} 
& \gtrsim \
\left\{
 \left( \frac{\chi_L^\dagger Y_E^{(2)}}{\Lambda^2}\right)_{ij} \,,
 \left( \frac{\chi_L^T X_E^{(2)}}{\Lambda^2}\right)_{ij} 
\right\} \,,
\nonumber \\[0.3em]
\left( \frac{Y_E^{(2)}}{\Lambda}\right)_{ij} 
& \gtrsim \
\left\{
 \left( \frac{\chi_L \, \xi_{E_R}^\dagger}{\Lambda^2}\right)_{ij} \,,
 \left( \frac{\phi_{1,2} \, X_E^{(2)}}{\Lambda^2}\right)_{ij} 
\right\} \,,
\label{inequal1}
\end{align}
where we have not quoted ``trivial'' inequalities involving $Z_L^{(2)}$ and $Z_E$.
The relations (\ref{inequal1}) are understood to 
hold order-of-magnitude-wise for a generic basis 
(i.e.\ where the off-diagonal entries of rotation matrices to the mass eigenbasis 
 are of natural size).

\subsection{Inverted hierarchy}

\begin{table}[t!pbh]
\begin{center} \small
 \begin{tabular}{|l || c | c | c | }
\hline
   & $E_R$ 
   & $({\cal U}_L^{(2)\dagger} L)$ 
   &  $(\Xi_L^\dagger L)$ 
\\
\hline 
$\bar E_R$ & $Z_E \sim(1,8)_0$ 
           & $Y_E^{(2)\dagger} \sim (2,3)_0$
           & $\xi_{E_R} \sim (1,3)_{-1}$ 
\\
$(\bar L \, {\cal U}_L^{(2)})$ 
           & $\bullet$ 
           & $\tilde g_\nu^{(2)} \sim (2,1)_0$
           & $\chi_L \sim (2,1)_{-1}$ 
\\
$(\bar L \, \Xi_L)$ 
           & $\bullet$ 
           & $\bullet$ 
           &  $ \sim (1,1)_0$ 
\\
 \hline
 \end{tabular}
\caption{\label{tab:nMFVinverted} 
Possible bi-linear combinations of fundamental fermion fields and
the associated spurion fields (inverted hierarchy,
$G_F'= SO(2)_L \times SU(3)_{E_R} \times U(1)_{L_3}$).}
\end{center}
\end{table}

The situation is slightly simpler in the case of inverted hierarchy, 
where the basic fermion fields with definite transformations under $G_F'$ are
\begin{align}
 & E_R \sim (1,3)_0 \,,  \qquad ({\cal U}_L^{(2)\dagger} L)  
 \sim (2,1)_0 \,, \qquad
  (\Xi_L^\dagger L) \sim (1,1)_{1} \,.
\end{align}
This implies the nMLFV spurion representations
in Table~\ref{tab:nMFVinverted},
which only introduces two new spurion fields, $\chi_L$ and $Z_E$.
In this case, the non-trivial inequality constraints for $\chi_L$ are obtained as
\begin{align}
 \left( \frac{\tilde g_\nu^{(2)}}{\Lambda}\right)_{ij} 
& \gtrsim \
 \left\{
\left( \frac{\chi_L \, \chi_L^\dagger + \chi_L^*  \chi_L^T}{\Lambda^2}\right)_{ij}  \,,
\left( \frac{ (Y_E^{(2)} \, Y_E^{(2)\dagger}) + (\cdots)^T - {\rm tr}(\cdots)}{\Lambda^2} \right)_{ij} 
\right\} \,,
\nonumber \\[0.3em]
\left( \frac{\xi_{E_R}^\dagger}{\Lambda}\right)_{ij} 
& \gtrsim \
 \left( \frac{\chi_L^T Y_E^{(2)}}{\Lambda^2}\right)_{ij} 
 \,, \qquad
\left( \frac{Y_E^{(2)}}{\Lambda}\right)_{ij} 
 \gtrsim \
 \left( \frac{\chi_L \, \xi_{E_R}^\dagger}{\Lambda^2}\right)_{ij} 
 \,.
\label{inequal2}
\end{align}

\subsection{Degeneracy}

Finally, for the case of degenerate neutrino masses, 
before the breaking of $G_F' \to G_F''$,
the nMLFV scheme reads\nopagebreak
\begin{center} \small
 \begin{tabular}{|l || c | c|   }
\hline 
   & $E_R \sim (1,3)$ 
   & $({\cal U}_L^{\dagger} L) \sim (3,1)$ 
\\
\hline 
$\bar E_R$ & $Z_E \sim(1,8)$ 
           & $Y_E^{\dagger} \sim (3,3)$
\\
$(\bar L \, {\cal U}_L)$ 
           &  $\bullet $ 
           & $\check g_\nu \sim (5,1) + Z_L \sim (3,1)$
\\
 \hline
 \end{tabular}
\end{center}\nopagebreak
which introduces the new spurions $Z_E$ and $Z_L$ 
with ``trivial'' inequality constraints.
Applying the nMLFV construction to the ET after the breaking of $G_F''$,
we obtain
\begin{center} \small
 \begin{tabular}{|l || c | c  | c |}
\hline 
   & $E_R \sim (1,3)_0$ 
   & $(\check{\cal U}_L^{(2)\dagger} L) \sim (2,1)_0$ 
   & $(\check\Xi_L^{\dagger} L) \sim (1,1)_1$ 
\\
\hline 
$\bar E_R$ & $Z_E \sim(1,8)_0$ 
           & $\check Y_E^{(2)\dagger} \sim (2,3)_0$
           & $\check\xi_{E_R} \sim (1,3)_1$
\\
$(\bar L \, \check{\cal U}_L^{(2)})$ 
           & $\bullet $ 
           & $\check g_\nu^{(2)} \sim (2,1)_0$
           & $\check \chi_L \sim (2,1)_1$
\\
$(\bar L \, \check\Xi_L)$ 
           & $\bullet $ 
           &  $\bullet $ 
           & $\sim (1,1)_0$
\\
 \hline
 \end{tabular}
\end{center}
which has a similar form as for the inverted hierarchy case, only the
$U(1)_{L_3}$ quantum numbers are replaced by $Z_2$ ones.

As a final remark, we should also point out that the nMFV framework would allow for
spurion fields that transform under both, the quark and the lepton-flavour
symmetry group, and could be a remnant of lepto-quark interactions which
typically appear in grand-unified theories. A detailed discussion of the
potential consequences and phenomenological constraints
is beyond the scope of this work.

\section{Summary}
\label{summary}
Non-linear realizations of flavour symmetry are advantageous in cases where
very distinct eigenvalues of Yukawa or Majorana mass matrices appear.
For the quarks, it is the large Yukawa coupling of the top which breaks the 
$SU(3)^3$ flavour symmetry down to $SU(3)\times SU(2)^2 \times U(1)$, where
the latter symmetry is only weakly broken by the remaining small Yukawa couplings.
In this paper we have used the same reasoning to discuss the flavour symmetries of 
leptons.  Here the hierarchy in the neutrino mass differences,
 $\Delta m^2_{\rm sol} \ll \Delta m^2_{\rm atm}$, can be used to construct
a parameterization of the effective Majorana mass matrix (entering
the dim-5 operator in a scenario with only left-handed neutrinos)
which reflects the non-linear realization of lepton flavour symmetry.

We have considered the various possible scenarios for neutrino-mass hierarchies, 
and for each case we have determined the residual symmetries, after the largest entries 
in the Majorana mass matrix have been identified. The remaining entries are 
parameterized in terms of Goldstone modes for the broken generators, and
spurion fields which eventually break the residual flavour symmetry.
Based on the minimal flavour violation hypothesis, 
we may then construct the flavour structure of possible New
Physics operators which mediate e.g.\ lepton-flavour violating decays of 
charged leptons. 
Within the same framework, we have also considered possible parameterizations 
of ``next-to-minimal lepton flavour violation'' 
along the lines proposed in \cite{Feldmann:2006jk}.

Besides offering a systematic model-independent\footnote{In this paper,
we only discussed a set-up with minimal field content, i.e.\ without right-handed
neutrinos and with only one Higgs doublet.} framework to discuss
deviations from the Standard Model in lepton-flavour violating processes,
our approach also provides some new perspectives on the flavour puzzle
within the Standard Model and beyond. In particular, it is interesting
to note that the different possible realizations of mass hierarchies
for neutrinos and charged leptons is unambiguously linked to different 
sequences of flavour symmetry breaking, $G_F \to G_F' \to G_F'' \to \ldots$ 
(A similar statement holds for the quark sector, which is going 
to be explored in a future publication).

\subsection*{Acknowledgements}
We thank Wolfgang Kilian and Werner Rodejohann for helpful discussions.
This work is partially supported by
the German Research Foundation (DFG, Contract 
No.~MA1187/10-1) and by the German Ministry of Research (BMBF,
Contract No.~05HT6PSA).

\end{document}